\documentclass[manuscript]{aastex}

\received{}
\revised{}
\accepted{}

\shorttitle{SMC star clusters}
\shortauthors{Piatti}

\begin{document}

\title{Identification of a new relatively old star cluster
 in the Small Magellanic Cloud}

\author{Andr\'es E. Piatti}
\affil{Instituto de Astronom\'{\i}a y F\'{\i}sica del Espacio, CC 67, Suc. 
28, 1428, Ciudad de Buenos Aires, Argentina}
\email{e-mail: andres@iafe.uba.ar}

\begin{abstract}

We present results on the age and metallicity estimates of the astonishingly unstudied 
SMC cluster ESO\,51-SC09, from CCD $BVI$ photometry obtained at the ESO NTT
with the EMMI attached. 
ESO\,51-SC09 turns out to be a relatively small cluster (FWHM = (10 $\pm$ 1) pc)
located $\sim$ 4$\degr$ northward from the galaxy center.
We report for the first time a mean cluster age of (7.0 $\pm$ 1.3) Gyr and a mean
cluster metallicity of [Fe/H] = (-1.00 $\pm$ 0.15) dex, concluding that 
ESO\,51-SC09 belongs to the group of the oldest SMC clusters.
We found that the cluster is projected onto a dominant field stellar population older 
(age $\sim$ 10-13 Gyr) and more metal-poor ([Fe/H] = -1.3$\pm$0.2 dex), so that the cluster could reach
its current location because of its orbital motion.

\end{abstract}

\keywords{techniques: photometric -- galaxies: individual (SMC) -- galaxies: star clusters.}

\section{Introduction}

As far as we are aware, ESO\,51-SC09  (RA = 00$^h$ 58$^m$ 57$^s$.96, Dec. = -68$\degr$ 54$\arcmin$ 
55$\arcsec$.7, J2000) is a cataloged star cluster of the
Small Magellanic Cloud \citep[SMC,][]{bietal08} which has remained unstudied until the present.
It is located in the outer disk of the SMC, at $\sim$ 4$\degr$ northward from the galaxy center as 
supposed to be at RA = 00$^h$ 52$^m$ 45$^s$, Dec. = −72$\degr$ 49$\arcmin$ 43$\arcsec$ (J2000) 
\citep{cetal01}. Its position should facilitate astrophysical studies, since reddening effects 
are at a minimum regime and the unavoidable field contamination does not represent a real constraint.
On the other hand, bearing in mind the enormous interest in identifying new relatively old/old clusters in the SMC 
\citep{detal10} and the appearance in the sky of ESO\,51-SC09 like a candidate old cluster \citep{l82}, it is astonishing 
that it has not been mentioned in the literature as a valuable target. 
According to the hierarchical star-formation scenario found by \citet{bb10}, the outer SMC disk would 
appear to be a genuine reservoir of old clusters. In fact, the oldest known cluster \citep[NGC\,121, age = 
10.6$\pm$0.5 Gyr,][]{detal01} is also placed in the outer disk, whereas Lindsay\,32 and 38 - two relatively old/old clusters 
\citep{petal01} - are in the ESO\,51-SC09's zone, at distances smaller than $\sim$ 1$\degr$. 

In this Letter we present for the first time age and metallicity estimates for ESO\,51-SC09.
The results show that this cluster belongs to the handful of oldest clusters in the SMC (age $\ge$ 5 Gyr). The impact of 
this finding would appear to be twofold: firstly, we actually found a new SMC cluster in the relatively old/old range.
Note that different campaigns have carried out until the present  searching for old star clusters in the 
SMC and, unfortunately, new candidates have not been
identified. These results would appear not only to show that the task of 
finding more old star clusters in the SMC is arduous, but also it would appear
a venture hardly to reach success. The amazing scarce amount of old SMC star
clusters results even more noticeable when comparing it with the 456 star
clusters cataloged in the SMC \citep{bb10}, thus representing $\approx$ 1\% of the
SMC star cluster population. Secondly, taking into account that \citet{p11a} predicted that 
we should expect to identify only one outer disk relatively old/old cluster 
not studied yet within those cataloged by \citet{bietal08}, and
ESO\,51-SC09 is uncovered to be a relatively old/old cluster, the SMC outer disk would appear not to be populated by any other 
unstudied old cluster.

This Letter is organized as follows: In Section 2 we describe the data collected, the reduction
procedures performed, and the subsequent photometry standardization. In Section 3 we deal with the
infamous cleaning process of the decontamination of field stars in the cluster Color-Magnitude
Diagram, while Section 4 is devoted to the estimation of the cluster age and metallicity. Finally,
Section 5 summarizes our results.

\section{Data collection and reduction}

Based on data obtained from the European Southern Observatory (ESO) Science Archive Facility under request number 33312, we
collected $BVI$ images centered on ESO\,51-SC09 along with observations of the  
SA\,113 and PG\,0231+51 standard fields \citep{l92} and calibration frames (zero, dark, sky-flat, dome-flat). 
The data were obtained at the ESO New Technology Telescope (NTT) with the ESO Multi-Mode Instrument (EMMI) 
attached (scale = 0.1665$\arcsec$/pixel). We obtained for ESO\,51-SC09 
exposures of 600 sec per filter under seeing conditions better than 0.$\arcsec$5, and with a mean airmass of 1.32. 
For the standard fields, the exposures varied from 10 up to 100 sec, depending on the filter and the airmass (airmass
range $\sim$ 1.1 - 2.2). 

The data reduction followed the procedures documented by the NOAO Deep Wide
Field Survey team (Jannuzi et al. 2003) and utilized the {\sc mscred} package in IRAF\footnote{IRAF is distributed by the 
National Optical Astronomy Observatories, which is operated by the Association of Universities for Research in Astronomy, 
Inc., under contract with the National Science Foundation.}. We processed a total of 274 images by
performing overscan, trimming corrections, bias subtraction, flattened all data images, etc., once the
calibration frames (zeros, darks, sky- and dome- flats, etc) were properly combined. Combined dark frames obtained from 
600 sec exposures showed count levels around zero. The final FOV of the images resulted to be $\sim$ 
4.7$\arcmin$$\times$4.7$\arcmin$ (see Fig. 1).

Nearly 15 independent magnitude measures of standard stars from the list of \citet{l92} were derived per 
filter  using the {\sc apphot} task within IRAF,
in order to secure the transformation from the instrumental to the $BVI$
standard system. The standard fields PG 0231+51 and SA\,113 contain
between 6 and 8 standard stars each distributed over an area similar to that of the
EMMI, so that we measured magnitudes of standard stars in each of its two
chips. The relationships between instrumental and standard magnitudes were obtained by
fitting the equations:

\begin{equation}
b = b_1 + V + (B-V) + b_2\times X_B + b_3\times (B-V)
\end{equation}

\begin{equation}
v = v_1 + V + v_2\times X_V + v_3\times (B-V)
\end{equation}

\begin{equation}
i = i_1 + V - (V-I) + i_2\times X_I + i_3\times (V-I)
\end{equation}

\noindent where $a_i$, $b_i$ and $c_i$ ($i$ = 1, 2, and 3) are the fitted coefficients, and $X$ represents the effective
airmass. Capital and lowercase letters represent standard and instrumental
magnitudes, respectively. We solved the transformation equations with the {\sc fitparams} task in IRAF for the 2 chips 
simultaneously, and found mean color terms of 0.028$\pm$0.009 in $b$, 0.006$\pm$0.014 in $v$ and
0.012$\pm$0.020 in $i$, and mean airmass coefficients of 0.243$\pm$0.008 in $b$, 0.141$\pm$0.013 in $v$ and
0.043$\pm$0.018 in $i$. The rms errors from 
the transformation to the  standard system were 0.010, 0.016 and 0.015 mag for $B$, $V$ and $I$, respectively, 
indicating an excellent photometric quality. 

The stellar photometry was performed using the star-finding and point-spread-function (PSF) fitting 
routines in the {\sc daophot/allstar} suite of programs \citep{setal90}. For each image, a quadratically varying 
PSF was derived by fitting $\sim$ 40 stars, once the neighbors were eliminated using a preliminary PSF
derived from the brightest, least contaminated $\sim$ 15 stars. Both groups of PSF 
stars were interactively selected. We then used the {\sc allstar} program to apply the resulting PSF to the 
identified stellar objects and to create a subtracted image which was used to find and measure magnitudes of 
additional fainter stars. This procedure was repeated three times for each frame. We examined the final subtracted
images and corroborated that all the stars in the cluster field were eliminated.
Finally, we standardized
the resulting instrumental magnitudes and combined all the independent measurements using the stand-alone {\sc daomatch} 
and {\sc daomaster} programs, kindly provided by Peter Stetson.  The final information  
consists of a running number per star, its $x$ and $y$ coordinates, the measured $V$ magnitudes and $B-V$ 
and $V-I$
colors, and the observational errors $\sigma(V)$, $\sigma(B-V)$ and $\sigma(V-I)$.
Only a portion of Table 1, which gives this information for a total of 1677 stars,
is shown here for guidance regarding its form and content. The whole content of Table 1 is 
available in the online version of the journal. Fig. 1 shows the $V$ magnitude and $B-V$ and $V-I$ color errors
as provided by {\sc daophot}. 

\section{Color-Magnitude Diagram cleaning}

Fig. 2 depicts the
$(B-V,V-I)$ diagram for all the measured stars
and the $(V,B-I)$ Color-Magnitude Diagrams (CMDs) for those stars located inside or outside a circle of
radius 160 pixels centered on the cluster. As can be
figured out,  this simple circular CMD extraction around the cluster center could 
lead to a wrong age, since this is obviously composed of stars of different stellar 
populations. Consequently, it is hardly possible to assess whether the whole extent of the observed populous
Main Sequence (MS) or the subgiant and red giant branches (SGB,RGB) trace the fiducial cluster features.

With the aim of cleaning the cluster CMDs from the unavoidable star field contamination we applied a 
procedure designed by \citet{pb12} which makes use of variable cells in the CMDs. The cells are adjusted in such a way 
that they result bigger in
CMD regions with a scarce number of field stars, and viceversa. This way, we reproduce the field CMD as closely as
possible on the cluster CMD. The method does not need to know whether a star is placed close to the cluster center
nor the cluster radial density profile to infer a membership probability. However, it takes into account the star field
density, since the more populous a star field the larger the number of stars subtracted from the cluster CMD. 
As a result, the intrinsic spatial star distribution is uncovered within the cluster region.
The method has shown to be able to eliminate stochastic effects in cluster CMDs
caused by the presence of isolated bright stars, as well as, to make a finer cleaning in the most populous CMD regions. 

For ESO\,51-SC09 we cleaned a circular region centered on the cluster with a radius twice as big as the Full-Width at
Half Maximum (FWHM) of its stellar density radial profile, as fitted by the IRAF {\sc ngaussfit} routine. 
As for the reference star fields, we used four different regions with the same area as for
the cluster. These circular field areas were placed toward the NE, NW, SE, and SW of 
the cluster farther than three times the estimated cluster's FWHM. This was done in order to take into account 
variations in the spatial density, magnitudes, and colors of field stars. 
Thus, we obtained four different cleaned CMDs. When comparing those CMDs, one may find 
stars that have kept not subtracted in most of the times, while other stars were subtracted in most of the program executions.
The different number of times that a star keeps not subtracted can then be converted into a measure of the probability of being a
fiducial feature of the cluster field. Thus, we were able to distinguish stellar populations projected onto
the cluster field that have a probability P $<$ 25\% of being a genuine cluster population, i.e., a typical 
foreground population; stars that could indistinguishably belong to the star field or to ESO\,51-SC09 (P = 50\%);
and stars that are predominantly found toward the cluster field (P $>$ 75\%) rather than in the star field
population. Fig. 3 shows the stars with chances of being a cluster feature higher than 75\%.

As can be seen, the cluster MS resulted successfully uncovered, especially its Turn Off (TO). In addition, some 
stars in the SGB and RGB, as well as red clump (RC) and blue and red horizontal branch 
stars also probably 
belong to the cluster stellar population. At a first glance, it appears that we are dealing with an  old SMC cluster.
Despite the presence of some interlopers and the absent of some cluster stars, this is the first time that 
a fiducial CMD is obtained for the cluster. From the cleaned cluster photometry, we fitted its stellar density
radial profile with a Gaussian function and obtained an intrinsic FWHM of (0.58 $\pm$ 0.06)$\arcmin$, which converts into
(10 $\pm$ 1) pc if a SMC distance of 60 kpc is adopted \citep{getal10} for the cluster.
On the other hand, the field stars observed along the line-of-sight toward ESO\,51-SC09 appear
to be composed by intermediate-age to old stars, although the dominant population is featured by a MS TO fainter (older)
than that of the cluster ($\Delta V_{\rm TO}$ $\sim$ 0.5 mag), by a noticeable populous SGB, and by
a RGB more metal-poor than that of ESO\,51-SC09, placed toward bluer colors.

\section{Age and metallicity estimates}

The age was first calculated by determining the difference in $V$ magnitude 
between the RC and the MS TO from the
cluster CMD (see Fig. 3) and then using the equation for the Morphological Age Index \citep[MAI][]{jp94} to obtain 
the age. As \citet{jp94} showed, MAI is well correllated with the logarithm of cluster ages, as determined by fitting
to theoretical isochrones. Note that 
this age measurement technique does not require absolute 
photometry and is independent of reddening as well. An additional advantage is that we do not need to go 
deep enough to see the extended cluster MS but only its MS TO.
The derived $\delta V$ difference resulted to be 2.50$\pm$0.15 mag;
its uncertainty $\sigma$($\delta V$) was estimated by considering the 
photometric errors at the RC and MS TO $V$ magnitudes (see Fig. 1) and the intrinsic dispersion 
in the CMD (Fig. 3). We adopted uncertainties five times larger than the photometric errors at the
RC and MS TO $V$ magnitudes. The computed cluster age turned out to be (8.3 $\pm$ 1.6) Gyr.
As can be seen, although the age error is slightly large, ESO\,51-SC09 is clearly 
a new discovered relatively old/old cluster in the SMC. 

Since \citet{getal97} showed that $\delta V$ is very well-correlated with
$\delta T_1$ (correlation coefficient = 0.993), we calculated the $\delta T_1$ index 
using their equation (3), and then derived an age estimate from their equation (4). 
The $\delta T_1$ index has proven to be a powerful tool to derive ages for star clusters older than 1
Gyr, independently of their metallicities \citep{betal98,petal11}. We obtained $\delta T_1$ = 
2.70 $\pm$ 0.15 mag, which in turn converts into an age of (6.5 $\pm$ 1.0) Gyr. This value is in
excellent agreement within the quoted uncertainties with that derived from $\delta V$. Note that
both luminosity differences were independently
calibrated in terms of cluster ages.

We finally fitted theoretical isochrones to the cluster CMD to get an additional
age estimate, by 
taking advantage of the theoretical isochrones
calculated with core overshooting by \citet{metal08}.  
We adopted chemical compositions 
from $Z$ = 0.001 to 0.004 in steps of 0.0005 for the isochrone sets which 
cover the metallicity range of most of the SMC relatively old/old clusters studied in detail so far \citep{detal10}.

We then selected a set of isochrones in steps of $\Delta$log$t$ = 0.05 and superimposed them on the cluster CMDs, 
once they were 
properly shifted by the corresponding $E(B-V)$ color excess and by the SMC apparent distance 
modulus. The estimation of the cluster reddening value was made by interpolating the extinction maps of 
\citet[][hereafter BH]{bh82}. 
They furnish us with a foreground $E(B-V)$ color excess equals to zero.
\citet[][hereafter SFD]{sfd98} obtained 
full-sky maps from 100-$\mu$m dust emission. 
We also used SFD's reddening maps to prove that when 
maximum and minimum extinction values are compared, the interstellar reddening is uniform across the 
observed field. We computed $E(B-V)_{SFD}$ color 
excesses for  a grid in the ({\it l},b) Galactic 
coordinate plane, with steps of $\Delta$({\it l},b) = (0.01$\degr$,0.01$\degr$) covering the whole 
observed field. We obtained 25 color excess values for the cluster field. Then, for the 
resulting $E(B-V)_{SFD}$ values, we built a histogram and calculated its center and FWHM. Since the 
FWHM value turned out to be quite small, we assumed that the interstellar absorption is uniform 
across the cluster field. Finally, we adopted $E(B-V)$ = 0.02 $\pm$ 0.01. As for the cluster distance modulus, 
we adopted the value of the SMC distance modulus 
$(m-M)_o$ = 18.90 $\pm$ 0.10 recently reported by \citet{getal10}. 
Considering 
BH reddening values for populous SMC clusters, \citet{cetal01} found the line-of-sight 
depth of the galaxy to be approximately 6 kpc. Then, bearing in mind that ESO\,51-SC09 could be placed 
in front of or behind the main body of the SMC, we concluded that the difference in apparent distance 
modulus could be as large as $\Delta(V-M_V)$ $\sim$ 0.2 mag, if a value of 60 kpc is adopted for the mean 
SMC distance. Given the fact that an uncertainty of 0.2-0.3 mag was estimated when adjusting the 
isochrones to the cluster CMD in magnitude, our simple assumption of adopting a unique value for 
the cluster  distance modulus should not dominate the error budget in our final result. In 
fact, when over-plotting the Zero Age Main Sequence on the observed cluster CMD, previously shifted by 
the $E(B-V)$ = 0.02 and $(m-M)_o$ = 18.90, an excellent match was generally found, especially
for the RC $V$ magnitude.

In the matching procedure, we used different isochrones for each metallicity 
level, ranging from slightly younger than the derived cluster age to slightly older. Finally, 
we adopted as the cluster age the one corresponding to the isochrone which best reproduced the cluster 
main features in the CMD. The presence of the RC, the SGB, the RGB in the cluster CMD made the fitting procedure 
easier, particularly in order to disentangle the cluster metallicity. The cluster age and metallicity
resulted to be $t$ = (6.3 $\pm$ 1.0) Gyr and [Fe/H] = (-1.00 $\pm$ 0.15) dex, respectively, where the
uncertainties come from different combinations of ($t$, [Fe/H]) pairs that reasonably adjust 
the observed dispersion in the cluster CMD. Fig. 3 shows the result 
of the fitting, where we plotted the isochrone of the adopted cluster age (log $t$ = 9.8)
and two additional isochrones bracketing the derived age (log $t$ = 9.7, 9.9) for $Z$ = 0.002.  
Finally, we averaged the three different age estimates and derived a mean age for ESO\,51-SC09 of
(7.0 $\pm$ 1.3) Gyr. This mean age places ESO\,51-SC09 within the oldest known clusters of the SMC, only
younger than NGC\,121 \citep[10.6 Gyr,][]{detal01}, HW\,42 \citep[9.3 Gyr,][]{p11b}, NGC\,361 \citep[8.1 Gyr,][]{metal98},
and Lindsay\,1 \citep[7.5 Gyr,][]{getal08}. We used the $\delta V$ index and the isochrone fitting to
the field star CMD to
estimate a mean age and metallicity of $\sim$ 10-13 Gyr and -1.3$\pm$0.2 dex, respectively, for the dominant star field
projected toward the cluster direction. The remarkable different ages and metallicities of ESO\,51-SC09 and 
the dominant field stellar population could be explained if we assume that the cluster was born in other part
of the galaxy and, because of its orbial motion, it is observed at the current location.

\section{Summary}

In this study we present for the first time CCD $BVI$ photometry
of stars in the field of the unstudied SMC cluster ESO\,51-SC09. 
The data were obtained at the ESO NTT with the EMMI attached
under high quality photometric conditions. We are confident
that the photometric data yield
accurate morphology and position of the main cluster features in the CMD.
To disentangle the cluster features from those belonging to its surrounding field, we applied
a subtraction procedure to statistically clean the cluster CMD from field star contamination. 
The method has shown to be able to eliminate stochastic effects in the cluster CMDs
caused by the presence of isolated bright stars, as well as, to make a finer cleaning in the 
most populous CMD regions. The FWHM of the genuine cluster stellar density radial profile
turned out to be  (0.58 $\pm$ 0.06)$\arcmin$, which converts into
(10 $\pm$ 1) pc if a SMC distance of 60 kpc is adopted.
Using the cleaned cluster ($V$,$B-V$) diagram, we estimated its age and metallicity 
using the $\delta V$ and $\delta T_1$ indices and fitting theoretical isochrones. 
The three different age estimates are in excellent agreement, resulting in a mean value
of (7.0 $\pm$ 1.3) Gyr. A metallicity of [Fe/H] = (-1.00 $\pm$ 0.15) dex was estimated
from the fitting of theoretical isochrones to the cluster CMDs.
We thus report that ESO\,51-SC09 belongs to the group of the oldest SMC clusters, only
younger (mean values) than NGC\,121, HW\,42, NGC\,361, and Linday\,1. The cluster is
placed in a region of the SMC where probably it has not been born, since the
mean age and metallicity of the dominant field stellar population is remarkable older
(age $\sim$ 10-13 Gyr) and more metal poor ([FeH] = -1.3$\pm$0.2 dex). The cluster could reach its
current location because of its orbital motion.

\acknowledgements
I greatly appreciate the suggestions raised by the
reviewer which helped me to improve the manuscript.
This work was partially supported by the Argentinian institutions CONICET and
Agencia Nacional de Promoci\'on Cient\'{\i}fica y Tecnol\'ogica (ANPCyT).

\begin{deluxetable}{lcccccccc}
\tabletypesize{\small}
\tablecaption{$BVI$ data of stars in the field of ESO\,51-SC09.}

\tablehead{\colhead{Star} & \colhead{$x$}  & \colhead{$y$} & \colhead{$V$} & 
\colhead{$\sigma$$(V)$}  & \colhead{$B-V$} & \colhead{$\sigma$$(B-V)$} & \colhead{$V-I$} & \colhead{$\sigma$$(V-I)$} \\
\colhead{}     & \colhead{(pixel)} & \colhead{(pixel)} & \colhead{(mag)} & 
\colhead{(mag)} & \colhead{(mag)} & \colhead{(mag)}  & \colhead{(mag)} & \colhead{(mag)}}

\startdata
-    &   -     &   -     &  -    &  -    &  -&  -   & - & -\\
     41 & 1168.79 & 156.49  & 20.345  &  0.009  &  0.752  &  0.015  &  0.866  &  0.029 \\
     42 & 1491.06 & 156.82  & 18.262  &  0.005  &  1.525  &  0.008  &  2.136  &  0.032 \\
     43 &  847.14 & 157.07  & 21.997  &  0.024  &  0.493  &  0.042  &  0.663  &  0.049 \\
-    &   -     &   -     &    -   &   -   & -&  -  &  -  &   \\
\enddata

\end{deluxetable}

\begin{figure}
\includegraphics[angle=0,width=17cm]{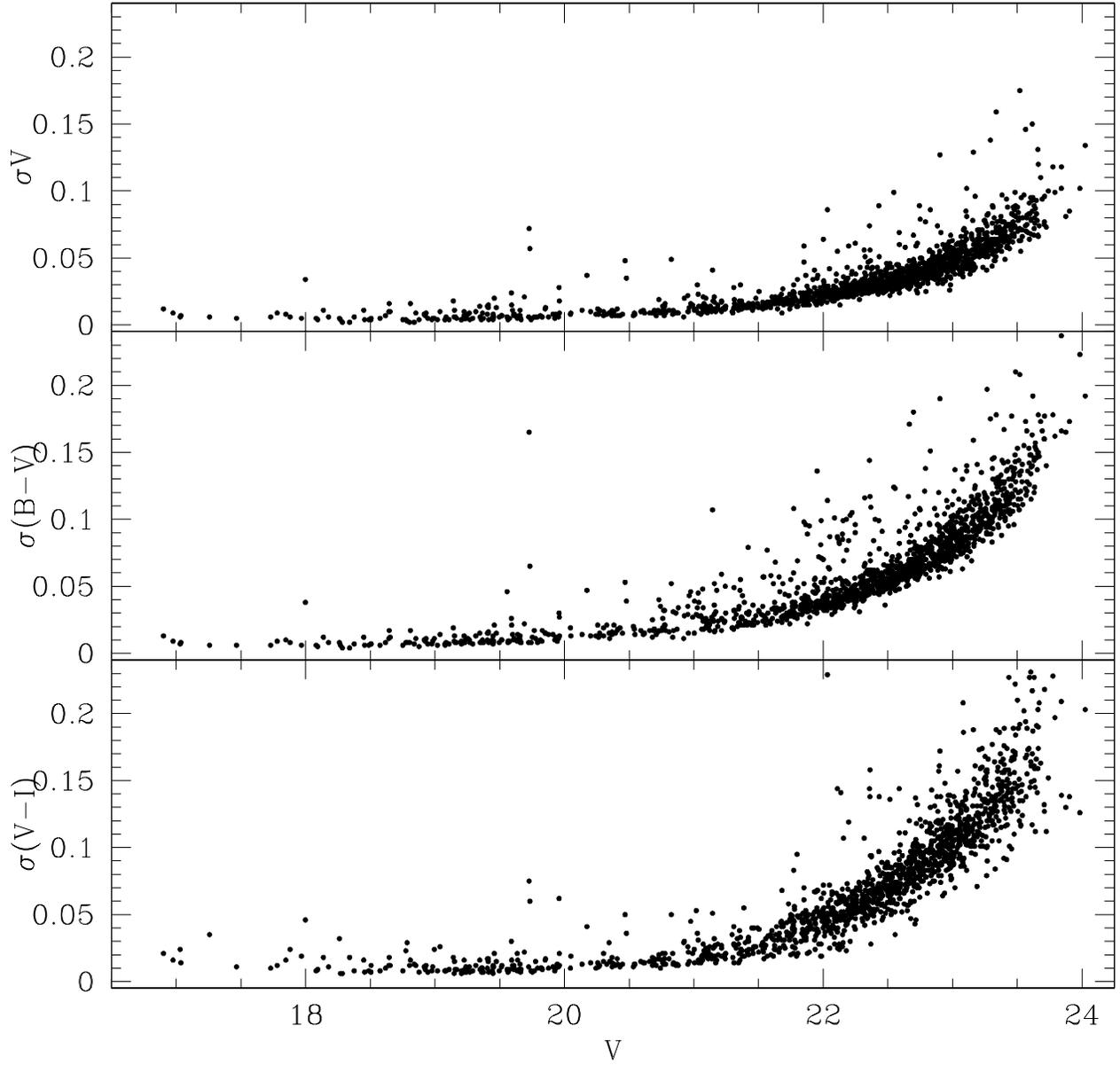}
\caption{Magnitude and color photometric errors as a function of $V$ for the stars measured in the field
of ESO\,51-SC09.}
\label{fig1}
\end{figure}

\begin{figure}
\includegraphics[angle=0,width=17cm]{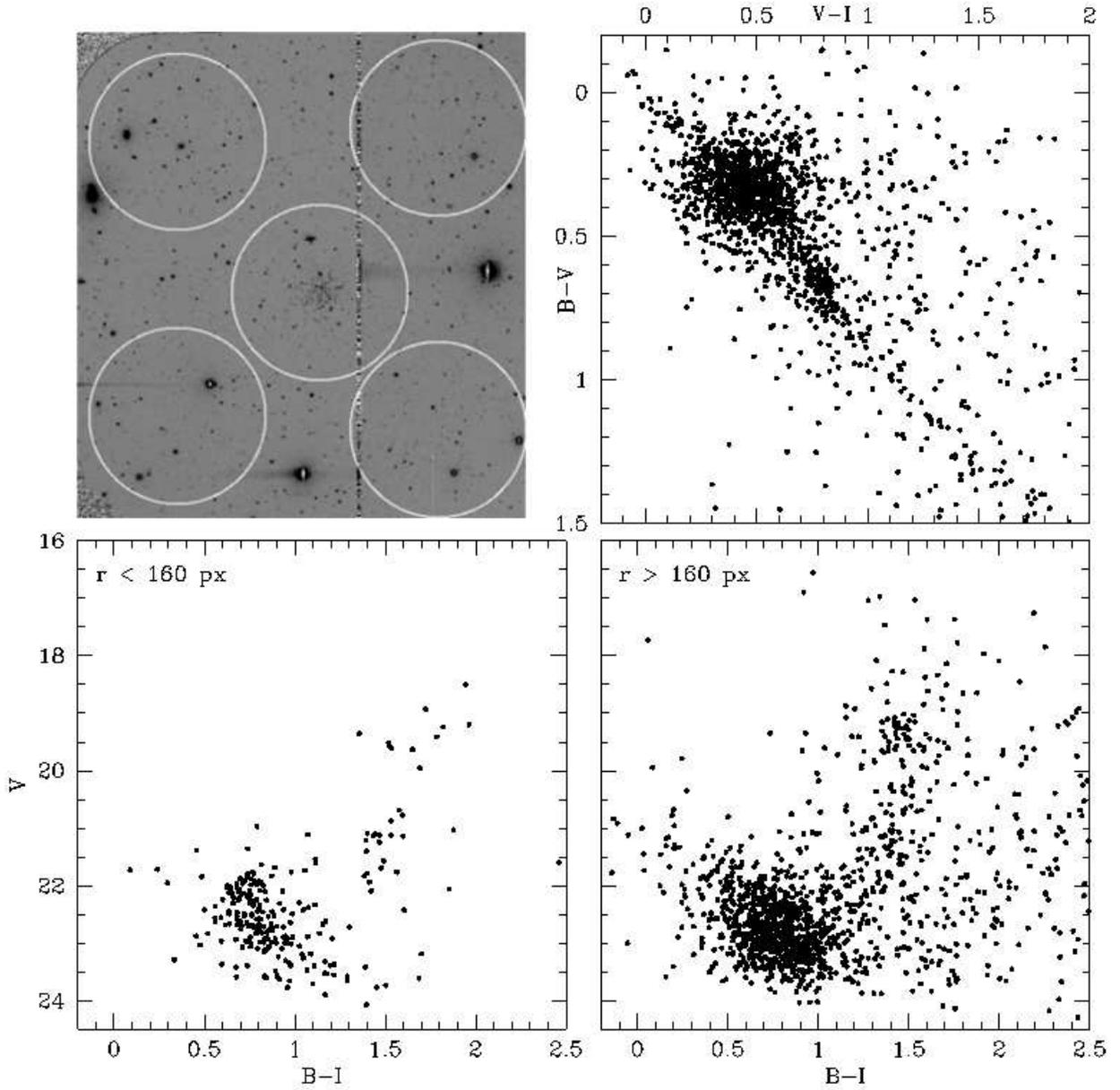}
\caption{
$V$ image with the cluster and four field circular regions over-plotted (upper left). The
radius of each circle is 320 pixels.
North is up and east is to the left. 
The ($B-V$,$V-I$) (upper-right)
and the ($V$,$B-I$) diagrams (bottom) for all the measured stars is also shown.}
\label{fig2}
\end{figure}

\begin{figure}
\includegraphics[angle=0,width=17cm]{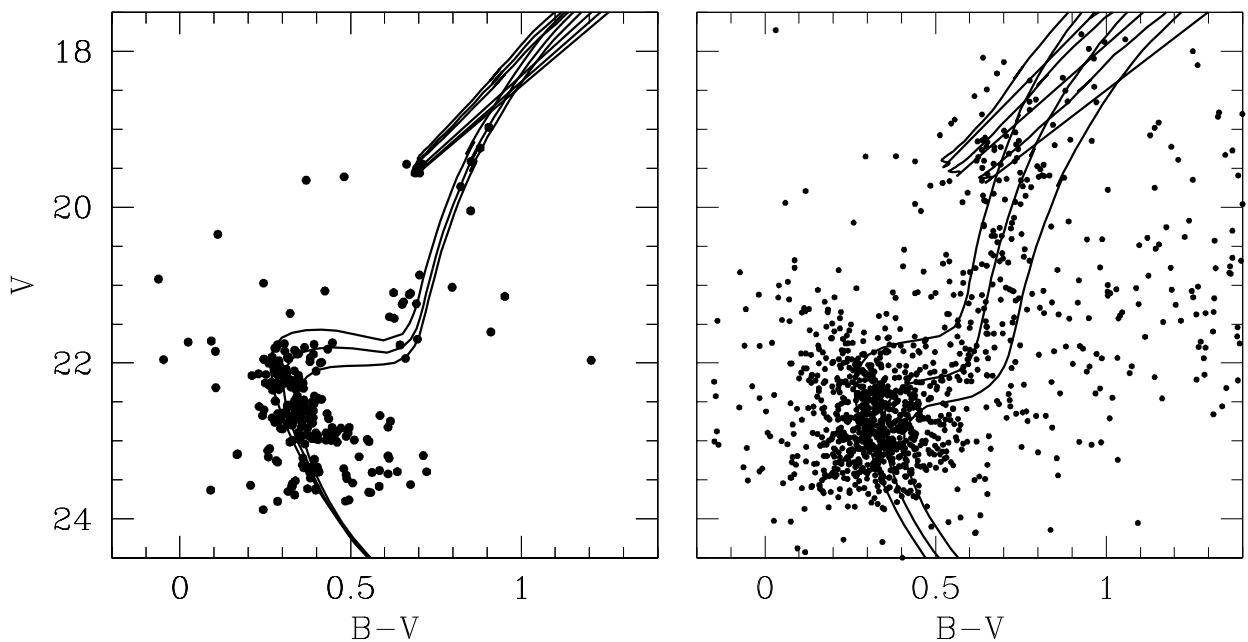}
\caption{{\it Left:} CMD of ESO\,51-SC09 with the Padova \citep{metal08} isochrones
for log$t$= 9.7, 9.8, and 9.9, and $Z$= 0.002. {\it Right:} Field CMD with isochrones
for log$t$= 9.9 (Z= 0.0005), 10.0 (Z= 0.001), and 10.1 (Z=0.002).}
\label{fig1}
\end{figure}

\end{document}